\documentstyle[mymulticol,pra,aps,amsfonts]{revtex}

\newcommand{\beq}{\begin{equation}}
\newcommand{\eeq}{\end{equation}}
\newcommand{\beqy}{\begin{eqnarray}}
\newcommand{\eeqy}{\end{eqnarray}}

\newtheorem{Postulate}{Postulate}
\newtheorem{Definition}{Definition}
\newtheorem{Lemma}{Lemma}
\newtheorem{Theorem}{Theorem}
\newtheorem{Corollary}{Corollary}

\newenvironment{Proof}{{\it Proof: \,}}{$\Box$ \vspace{0.3cm}}

\newenvironment{Definition*}{{\bf Definition}}{}

\makeatletter
\def\@beginTheorem#1#2{\trivlist \item[\hskip \labelsep{\bf #1\ #2}]}
\def\@opargbegintheorem#1#2#3{ \trivlist
      \item[\hskip \labelsep{\bf #1\ #2\ (#3)}]}
\makeatother

\makeatletter
\def\@beginLemma#1#2{\trivlist \item[\hskip \labelsep{\bf #1\ #2}]}
\def\@opargbeginLemma#1#2#3{ \trivlist
      \item[\hski
 Hence we have the same statements
about the increase of the supports for increasing depth, where
the local transformations are not counted for the depth.
p \labelsep{\bf #1\ #2\ (#3)}]}
\makeatother

\makeatletter
\def\@beginDefinition#1#2{\trivlist \item[\hskip \labelsep{\bf #1\ #2}]}
\def\@opargbeginDefinition#1#2#3{ \trivlist
      \item[\hskip \labelsep{\bf #1\ #2\ (#3)}]}
\makeatother

\makeatletter
\def\@beginCorollary#1#2{\trivlist \item[\hskip \labelsep{\bf #1\ #2}]}
\def\@opargbeginCorollary#1#2#3{ \trivlist
      \item[\hskip \labelsep{\bf #1\ #2\ (#3)}]}
\makeatother

\makeatletter
\def\@beginExample#1#2{\trivlist \item[\hskip \labelsep{\bf #1\ #2}]}
\def\@opargbeginExample#1#2#3{ \trivlist
      \item[\hskip \labelsep{\bf #1\ #2\ (#3)}]}
\makeatother

\def\C{{\mathbb{C}}}

\def\N{{\mathbb{N}}}

\newcommand{\cH}{{\cal H}}

\title{Remark on multi-particle observables and entangled 
 states with constant complexity}

\author{D. Janzing\thanks{Electronic address: janzing@ira.uka.de}  and 
Th. Beth}
\address{Institut f\"ur Algorithmen und Kognitive Systeme, Am Fasanengarten 3a,
    D--76\,128 Karlsruhe, Germany}

\begin{document}
\maketitle

\begin{abstract}
We show that
every density matrix of an $n$-particle system 
 prepared by a quantum network  of constant depth is asymptotically commuting
with the  mean-field observables.
  We introduce certain pairs of  hypersurfaces in the space
of density matrices and give lower bounds for the depth of a network
which prepares states lying outside those pairs.
The measurement of an observable which is not
asymptotically commuting with the mean-field observables
requires a network of depth in the order of $\log n$, if one   
demands  the measurement to {\it project} the state into the eigenspace
of the measured observable.
\end{abstract}

\begin{multicols}{2}

\section{Introduction}
In the standard formulation of quantum mechanics the states
are the positive operators with trace 1 on the system's Hilbert space\footnote{Here we ignore the case
of a quantum system with superselection rules.}.
The observables are the self-adjoint operators  and their eigenvalues
represent the possible measurement outcomes.
The question of how to design a {\it preparation procedure}
corresponding to a given trace-one operator or a {\it measurement
procedure} corresponding to a self-adjoint operator
remained unclear for many decades. Why should
 such preparation or measurement procedures
exist at all? Their existence  seemed to be just a common {\it belief} 
of the community of physicists.
Being aware of the abundance of states and observables
corresponding to a many-particle  quantum system, 
one might question this postulate.
Since the old problem of Schr\"odinger's Cat
is essentially the question of the set of
states and observables of macroscopic systems \cite{Om,GJKKSZ}, one should
even accept this question as a problem of {\it philosophical} relevance.

However, in the context of quantum  computing research and 
due to the recent experimental and theoretical  progress 
 in quantum optics, this question became a serious subject
of research (e.g. \cite{DMPS,RZBB,Sh,CR,SF,BBDJM,Ki,Ol}).
The connection of the problem described above to the subject 
of quantum computing research can roughly be sketched as
follows: 
 Define an {\it ideal quantum computer} as a quantum system
fulfilling the following conditions:
\begin{enumerate}
\item There is a physical procedure preparing one pure state 
$|\psi\rangle \in \cH$, where $\cH$ is the system's Hilbert space.
\item There is a set of (`basic') unitary transformations acting on $\cH$ which can be
 implemented by a physical process and which are universal in the sense, that
any unitary transformation can approximately be obtained by applying a sequence of basic transformations.
\item There is a read-out mechanism given by the measurement
of  one non-degenerated observable
$a$ acting on $\cH$.
\end{enumerate}
In an ideal quantum  computer, every pure state can approximately 
be prepared by performing the appropriate unitary transformations after having 
prepared the state $|\psi\rangle$.
Due to the evident operational meaning of convex combination on the set
of density matrices, one has preparation procedures for every density
matrix on $\cH$.
Analogously, one can find a measurement procedure for any self-adjoint operator
$b$ by writing it as $b=uf(a)u^*$ for an appropriate unitary operator $u$
and an appropriate function $f$.
Then $b$ can be measured by the procedure: `implement $u$, 
measure\footnote{Note that this measurement procedure does not project
the state vector into the eigenspaces of $b$, it only reproduces the 
correct probabilities for the measurement outcomes. Stronger senses
of measurement procedures will be considered below.}  $a$
afterwards, and apply the function $f$ to the result.'

In our opinion this framework allows a formulation of
 the problem of Schr\"odinger's Cat
in a way which is more explicit than it has ever been 
before: On the one hand, one has strong evidence
for the belief that  a system being composed from many particles
has {\it quantum and classical} aspects\cite{Se}, on the other hand, if the system
fulfills the axioms of the ideal quantum  computer, {\it there is no}
non-trivial observable which is compatible with all the other ones.

But the oversimplified answer `we can measure and prepare everything'
 is a result of idealized assumptions.
One may even state that these assumptions ignore in some sense the laws
of thermodynamics: If $\cH$ is the state space of a many particle system,
the preparation of a pure state $|\psi\rangle$ would even violate the
Third Law. Furthermore, fundamental bounds on the preparation
of states and the measurability might stem from the laws of quantum mechanics
itself \cite{Sh}.

Among other things, a serious analysis has to take into account the following
objections to the idealized assumptions:
\begin{enumerate}
\item {\bf The problem of complexity:}
Generically,  the set of basic transformations will be quite small
compared to the whole set of unitary operations.
Accordingly, the number of basic transformations needed for
the implementation of  a generic one will grow rather fast with  the size 
of the state space $\cH$.
 The statement `every unitary transformation can be implemented
 {\it in principle}' becomes doubtful for particle numbers
of macroscopic order. 
\item {\bf The problem of reliability:} We can neither expect
that the  preparation procedure leads to a pure initial state 
 nor that
any realizable operation acts on the density matrix 
like a conjugation with a unitary operator.
Taking into account a finite error  probability  
 it will rather be 
another completely positive map.
If a preparation or measurement procedure relies on a rather complex
iteration of basic operations, it is a non-trivial problem
to determine whether the procedure is sensitive to
errors during  the implementation of the  basic operations.
In \cite{JB}, for instance, we have shown that the preparation of states
showing {\it quantum} uncertainty on the macroscopic level 
in a sense explained below  would  require quite small error probabilities 
for the basic operations.
\end{enumerate}

Of course one could consider these statements
rather as statements about `practical' problems of realization
of quantum computers  and question 
its relevance with respect to the fundamental question described
 in the beginning. But one should not ignore that limitations
in accuracy of processes might be deeply connected with
thermodynamics. In \cite{JWZGB}, for instance, we analyzed in
 which sense the resource requirements
 for preparing an (approximate) pure state 
grow for increasing reliability.
Whether or not such  thermodynamic statements on resource requirements 
puts fundamental restrictions to the set of accessible states	
and measurable observables might be answered by the future.

Having in mind the problem of Schr\"odinger's Cat, we restrict our attention to
mean-field observables (see \cite{DW}) of many particle systems, which are the
best candidates for constituting the classical aspect of the system since
they represent at least one part of the {\it macroscopic} level.
Then it is natural to ask the following two questions:
\begin{enumerate}
\item
How difficult is it to prepare a state showing {\it large}
quantum uncertainty on the macroscopic level?
\item
How difficult is it to measure an observable which is {\it strongly} 
incompatible with a macroscopic one?
\end{enumerate} 

Both question cannot really be discussed separately:
If one wants to distinguish whether a strong variation of measurement
outcomes for any observable stems mainly from a {\it classical} statistical
 mixture
 or rather from a {\it quantum} superposition one has to measure an 
observable which is incompatible with the first one.
On the other hand, a measurement of an observable being incompatible
with the first one, could lead to a {\it quantum} superposition of macroscopically
distinct states.

The question `how difficult is it to prepare a superposition of macroscopically distinct states?' can be understood in various ways:
It might be the question for the {\it reliability} of the basic operations 
which would be required
for such preparations or measurements (this is partly answered in \cite{JB}),
it can also be interpreted as the question of {\it complexity}
of the required algorithm. This question will be focussed on here.

We will prove lower bounds for the depth  of a quantum network producing
macroscopic superpositions. 
Since these bounds are not strong,
 we do not claim this would be a serious restriction
for the realizability, we merely consider it as
an important part of pure research to determine such bounds.
Furthermore it gives insights in the structure of multi-particle entanglement
since we show lower bound for the complexity of
the same  class of highly entangled states which has been 
investigated in \cite{JB}.
Here a quantum network is defined to be a transformation being composed
by a  sequence of bilocal transformations acting on only two tensor components
at once.
On the one hand, this is a natural assumption from the
computer scientist's point of view, since it corresponds to the
decomposition of logical networks in 2-bit gates in the theory of classical
 electronic devices.
On the other hand, even from the physicist's point of view,
it is less artificial than it may seem at first sight:
If one assumes the qubits to correspond to particles, bipartite
interaction represent the structure of the fundamental forces of nature.

\section{Macrorealism and large quantum computers}

Let $\cH:=\otimes_{i=1}^n \C^{\,l}$ for arbitrary $l\in \N$ be
the Hilbert space of $n$ identical  quantum systems.
Let $(a_i)_{i\leq n}$ be a family of self-adjoint operators
with  the property, that every $a_i$ acts on the $i$-th component of the
n-fold tensor product only, i.e.
\[
a_i=\underbrace{1\otimes 1\dots 1}_{i-1}\otimes c \otimes \underbrace{1\otimes  \dots \otimes 1}_{n-i}
\]
with an arbitrary self-adjoint $c$ acting on $\C^l$.
Let $\|c\|\leq 1/2$ where $\|.\|$ denotes the operator norm
defined by $\|c\|:=\max_{|\psi\rangle}\|c|\psi\rangle\|/\|\,|\psi\rangle\|$.
Then we call
\[
\overline{a}:=\frac{1}{n}\sum_i a_i
\]
the corresponding {\it averaging observable}.
In solid state physics,
the algebra generated by them is usually referred to as
the algebra of {\it mean-field} observables \cite{DW}.
We claim, that they are some of the best candidates
 for constituting the {\it macroscopic} level
of the many-particle system. This might be made plausible by
taking the following example:

In a system consisting of $n$ spin-1/2 particles, the mean-magnetization is
described by the observables
\[
(1/2)\overline{\sigma}_x, \,\, (1/2)\overline{\sigma}_y, \,\, (1/2)\overline{\sigma}_z
\]
where $\overline{\sigma}_i$ with $i=x,y,z$ is defined as 
\[
\overline{\sigma}_i:=\sum_j \sigma_i^{(j)} 
\]
and $\sigma_i^{(j)}$ is the Pauli matrix $\sigma_i$ acting on the j-th 
tensor component.
Obviously the magnetization represents a physical quantity 
which has strong direct evidence in every day life.

Let $\|.\|_{tr}$ denote the trace norm of any matrix. It is defined
by $\|a\|_{tr}:=tr(\sqrt{a^\dagger a})$.
In \cite{JB} we argued, that a large value of the expression
\[
e_{\rho}:=\max_{\overline{a},\|b\|\leq 1}|tr(\rho [\overline{a},b])|=\max_{\overline{a}}\| [\overline{a},\rho]\|_{tr}
\]
indicates {\it quantum}  uncertainty
on the macroscopic level\footnote{Note that we take a definition
 slightly differing from that one introduced in \cite{JB}:
 Since we are dealing with the space
$\C^l$ on each site instead of restricting the proofs to the case
of qubits, i.e. $l=2$, it does not seem to be appropriate to restrict the
one-site-observables to projections.}.  
 Here {\it large} means that the size of the
expression is rather in the order of 1 than of $1/\sqrt{n}$, the latter 
is the case for separable states \cite{JB}.  
If one defines a family of hypersurfaces in the set of density matrices by
\[
H_{\overline{a},b,r}:=\{ \rho \,|\, tr(\rho [\overline{a},b])\}=r\}
\]
we find that for every $r\geq 0$
\[
e_\rho\leq r
\]
is equivalent to 
\[
|tr(\rho[\overline{a},b])|\leq r  \,\, \forall b \hbox { with }\|b\|\leq 1, \,
 \forall \overline{a} \,.
\]
Hence the convex set $\{\rho \, |\, e_\rho \leq r\}$ is enclosed by
the family of  pairs of  hypersurfaces
\[
(H_{\overline{a},b,\pm r})_{\overline{a},b}\, ,
\]
where $\overline{a}$ is any averaging observable and $b$ and $\|b\|\leq 1$.
In the following we will prove lower bounds for the depth of a quantum network
required for crossing these hypersurfaces.

To formalize the term `quantum network' we introduce the
following terminology:

\begin{Definition}
A {\bf bilocal} or {\bf local} map on a quantum system with
Hilbert space 
\[
\cH:=(\C^l)^{\otimes n} \,\, \hbox{ with } \,\, l,n \in \N 
\]
is a completely trace preserving map on the set of density matrices on $\cH$
acting trivially on each tensor component except two or one, respectively,
 i.e., it is given by
a canonical  embedding of a completely positive trace preserving map
acting on the density matrices on the Hilbert space $(\C^{\,l})^{\otimes 2}$
or $\C^{\,l}$, respectively.
The tensor components $\{i,j\}\subset \{1,\dots,n\}$ or
 $\{i\}\subset \{1,\dots,n\}$ on which the map 
acts nontrivially is called
its {\bf support}.
\end{Definition}

\begin{Definition}
A {\bf step} of a quantum network is a set $S:=\{G_1,\dots,G_m\}$
 of bilocal or local maps 
with mutually disjoint supports. Write $S(\rho)$ for 
$(G_1\circ\dots G_m)(\rho)$.

A {\bf  quantum network} of depth $k$ is a sequence $A:=(S_1,\dots,S_k)$
 of $k$ steps.
Write $A(\rho)$ 
for $(S_k\circ \dots \circ  S_1)(\rho)$.
By duality, $A$ defines a completely positive unital\footnote{A map $G$ on the observables is called
{\it unital} if one has $G(1)=1$ for the trivial observable $1$. 
One can show \cite{Pa} that completely positive
and unital implies that $G$ is norm decreasing, i.e.,
 $\|G(a)\|\leq \|a\|$ for every operator $a$.}
map $A^*$
on the set of observables by
\[
tr(A^*(a) \rho)=tr(aA(\rho))\,.
\] 
\end{Definition}

Furthermore we define:

\begin{Definition}
The {\bf support} of an observable $a$ is the set $S\subset \{1,\dots,n\}$
 of tensor components on which the operator $a$ acts nontrivially.
\end{Definition}
We observe:

\begin{Lemma}
Let $a$ be an observable with support of size $l$. Let $A$ be a quantum 
network  of depth $k$. Then the support of $A^*(a)$ is $l\cdot 2^k$ at most.
\end{Lemma} 

\begin{Proof}
By easy induction: Every step of $A$ can double the size of the support
at most.
\end{Proof}

Now we are able to prove one of our main statements:

\begin{Theorem}\label{Main}
Let $\rho$ be a separable state. Let $A$ be a quantum network of depth
$k$. Then we have 
\[
e_{A(\rho)}\leq \sqrt{\frac{2}{n}} 2^k
\]  
\end{Theorem}

\begin{Proof}
By convexity arguments
it is sufficient to proof the theorem for the case that $\rho$ is a
product state.
By Lemma 1 in \cite{JB} it is sufficient to prove
\[
\sqrt{tr(\overline{a}^2 A(\rho)) - (tr(\overline{a}A(\rho)))^2}\leq
 \frac{1}{\sqrt{2n}} 2^k
\]
for every $\overline{a}$.
We have:
\begin{eqnarray}\label{Varianz}
&&tr(\overline{a}^2 A(\rho))-(tr(\overline{a}A(\rho)))^2\\ \nonumber
&=&\frac{1}{n^2}\sum_{ij} (tr(A^*(a_i)A^*(a_j) \rho)-tr(A^*(a_i)\rho)\,
tr(A^*(a_j)\rho))\\  \nonumber
&=:&\frac{1}{n^2}\sum_{ij} g_{ij}\,.
\end{eqnarray}
Let $X_i$ be the support of $A^*(a_i)$. Since $\|A^*(a_i)\|\leq \|a_i\|\leq
 1/2$
we have $|g_{ij}|\leq 1/2$. Since $\rho$ is a product state we have
$g_{ij}=0$ for all those pairs $i,j$ with $X_i\cap X_j=\emptyset$.
Now we just have to estimate the number of those areas $X_j$ which intersect
a given area $X_i$. Firstly we show (by induction over $k$) that
for every site $y\in \{1,\dots,n\}$ there are at most $2^k$ areas $X_j$
with $y\in X_j$: For $k=0$ the statement is obvious. Let $B^*$ 
the dual map corresponding to  
 a quantum network $B$ obtained from $A$ by adding one more step.
Let $W_i$ be the supports of $B^*(a_i)$.
Assume that $y\in X_j$ for at most $2^k$ areas $X_j$.
Let the additional transformation step act on the pair $(y,x)$ where
 $x\in\{1,\dots,n\}$ is 
an arbitrary site with $x\neq y$. Then we have $y\in W_j$ only if $y\in X_j$
or $x\in X_j$. Since both sites are contained in at most $2^k$ areas $X_j$
we see that $y$ can be an element of at most $2^{k+1}$ areas $W_j$.
This completes the induction.

Since every area $X_i$ has size $2^k$ at most, for every $X_i$ there are $2^k2^k$ areas
$X_j$ at most with nonempty intersection. Hence $g_{ij}\neq 0$ for
$4^k$ pairs $(i,j)$ at most.\footnote{Actually the latter statement is
essentially  the dual formulation of Lemma 8 in \cite{AKN}. We preferred the
formulation on the set of observables since it turns out to be appropriate for
the proof of a theorem below.}
Hence the term (\ref{Varianz}) is less or equal to
$4^k/(2n)$.
\end{Proof}

The theorem can be rephrased in terms of the hypersurfaces
introduced above:

\begin{Corollary}
Let $r>0$. Let $b$ with $\|b\|\leq 1$ 
arbitrary and $\overline{a}$ an averaging observable.
Let $\rho$ be a separable initial state
and $A$ a quantum network such that
$\rho$ and $A(\rho)$ lie on different sides of the hypersurface
\[
H_{\overline{a},b,r}.
\] 
Then the depth of $A$ is at least
\[
\frac{\ln r-\ln (\sqrt{2/n})}{\ln 2}\,.
\]
\end{Corollary}

Note that the bound given in Theorem \ref{Main} is not too far from being 
tight:
Take an $n$-qubit quantum computer
with $n=2^k$ starting with the initial  state 
\[
\frac{1}{\sqrt{2}}(|0\rangle +|1\rangle)|0\rangle^{\otimes (n-1)}\,.
\]
Perform a controlled-not with  qubit 1 as  control-qubit qubit 2
 as  target. Than perform two controlled-not with qubit 1 and 2
as control-qubit and 3 and 4 as targets. In the r-th step one takes the 
qubits $1,\dots,2^{r-1}$  as control qubits and the qubits 
$2^{r-1}+1,\dots 2^r$ as targets. After $k$ steps one has the cat state
\[
\frac{1}{\sqrt{2}}|0\rangle^{\otimes n} +|1\rangle^{\otimes n}\,,
\]
i.e., a state with $e_\rho=1$ (see \cite{JB}: Despite the slight modification
in the definition of $e_\rho$ one can adopt the proof for the equation 
$e_\rho=1$ given  therein.).

Since we have answered the question of the complexity of a quantum network
producing states which show large quantum uncertainty on the macroscopic level
we shall focus on the second question
of the complexity of measurement procedures for observables which are strongly
incompatible with the averaging  ones.

Firstly we should make clear, what we mean by `measurement procedure
for an observable $a$', since this will turn out to be essential:

\begin{Definition}
Let $a$ be a self-adjoint operator acting on a quantum system's
Hilbert space of  finite dimension.
 Let $a=\sum_{i\leq j} \lambda_i P_i$ be its spectral decomposition.
\begin{itemize}
\item {\bf weak sense of measurement:}

A  procedure is said to be a {\bf measurement procedure for $a$}
if it has
 outcomes $\lambda_1,\dots,\lambda_j$
and  the outcome $\lambda_i$ has the probability
$tr(\rho P_i)$ for a system prepared  in the state $\rho$.

\item {\bf strong sense of measurement:}

If the procedure changes the state in such a way that one obtains the state
\[
\frac{P_i\rho P_i}{tr(\rho P_i)}
\]
in case of the result `$\lambda_i$' we call the procedure a
{\bf `von-Neumann-measurement'}. 
Note that this assumption is much stronger than the requirement that the state
vector lies  in the eigenspace of the measured eigenvalue after the measurement. It even has to be {\bf projected}.\footnote{In the terminology of \cite{Ja}
this is  an {\it ideal measurement} and corresponding to
 every spectral projection 
one has realized a {\it passive filter}.}
\end{itemize}

\end{Definition} 
	
Now we should make clear which kind of observables we assume to be measurable 
{\it directly}:

\begin{Postulate}
In the {\bf  weak sense}, one can measure every observable which is a function of
an observable $a$ of the form
\[
a=\otimes_i a_i
\]
where every $a_i$ is a self-adjoint operator acting
 on the $i$-th component of the tensor product.
\end{Postulate}

\begin{Postulate}
In the {\bf strong sense}  one can only measure those
 observables 
$a$ directly which have the property  that every spectral
 projection $P_i$ of $a$ is of the form
\[
P_i=\otimes Q_j^{(i)}\, ,
\]
where $Q_j^{(i)}$ is an orthogonal projection acting on the  $i$-th tensor 
component.
\end{Postulate}

These postulates assume, that a quite natural way of a  measurement 
of a many-particle system is given by measuring some or all the  particles
separately. The set of possible outcomes  is then given by the cartesian
product of the sets of possible outcomes of the one-particle measurements.
Of course it would also be  natural to assume that one single
 measurement apparatus interacts with many particles in the same way.
Then one would obtain rather complicated measurements as basic ones.
The question `which observable can be measured in the {\it most direct}
way' is hard to answer and it is not even clear how to give 
a definite meaning to the term `most direct'. 
Our postulates  should merely be considered as a first attempt to formalize
the intuitive evidence for the fact that many-particle systems
have an abundance of observables which seem to  require rather sophisticated
measurement procedures (in case they exist at all).

To illustrate  that the difference between the weak  and the strong sense
of measurement is essential we take the observable
\[
\otimes_i \sigma_x^{(i)} \, ,
\]
which
can be measured directly in the weak sense by measuring every qubit
in the eigenbasis of $\sigma_x$.
This observable  is strongly
incompatible with the averaging observable
\[
\overline{\sigma}_z
\]
in the sense that one has
\[
\|[\overline{\sigma}_z,\otimes_i \sigma_x^{(i)}]\| = 2\,.
\]
Easy calculation shows, that the observable
\[
\otimes_i \sigma_x^{(i)}
\]
is one of the best  for distinguishing
between the macroscopic {\it coherent} superposition
\[
\frac{1}{2}(|0\rangle^{\otimes n} \langle 0|^{\otimes n} + |1\rangle^{\otimes n} \langle 0|^{\otimes n}+|0\rangle^{\otimes n} \langle 1|^{\otimes n}+|1\rangle^{\otimes n} \langle 1|^{\otimes n}) 
\]
and the corresponding {\it mixture} of macroscopic distinct states:
\[
\frac{1}{2}(|0\rangle^{\otimes n} \langle 0|^{\otimes n}+|1\rangle^{\otimes n} \langle 1|^{\otimes n})\,.
\]
It is easy to see that our way of measuring $\otimes_i \sigma_x^{(i)}$
is far away from being a von-Neumann-measurement:
Our procedure discriminates $2^n$ different measurement outcomes 
in order to
measure an observable with only two different eigenvalues.
A measurement in the strong sense can be designed as follows:

\begin{enumerate}
\item Perform a unitary transformation $u$ such that 
\[
u^\dagger (\otimes_i \sigma^{(i)}_x)  u= 1\otimes 1 \otimes \dots \otimes 1\otimes \sigma_x\,.
\] 
\item Measure the observable
\[   
1\otimes 1 \otimes \dots \otimes 1\otimes \sigma_x \, .
\]
\item Perform the transformation $u^\dagger$\,.
\end{enumerate}

Essentially, these methods for measuring
a highly degenerated observable without destroying 
`too much coherence' are necessary for quantum error correction since 
the computation of the error syndrome \cite{CRSS} must not destroy 
 the encoded quantum information.

But why is the difference between the two ways of measurement so important
for {\it our} main question?
-- Because their effects on the states
with respect to the problem of
Schr\"odinger's Cat 
 are in some sense 
even {\it complementary}: While the first one {\it destroys} macroscopic
superpositions, the second one can {\it prepare} them. 
This can be seen as follows:
Perform the first measurement procedure. Depending on the measurement
 outcomes, it will
produce one of the $2^n$ states
\[
\otimes_i |\pm x_i\rangle ,
\]
where $|\pm x_i\rangle$ denotes the eigenstate of the Pauli matrix $\sigma_x$
with eigenvalue $+1$ or $-1$ at the i-th qubit.
Note that for every initial state the measurement produces  a product state.
  On the other hand,
take the initial state
\[
|0\rangle^{\otimes n}
\]
 and  perform the second 
procedure.
Since the projectors on the eigenspaces of $\otimes_i \sigma_x^{(i)}$
are given by  
$P_-:=\frac{1}{2}(\otimes_i \sigma^{(i)}_x +1)$
and $P_+:=\frac{1}{2}(1- \otimes_i \sigma^{(i)}_x)$  we obtain
\[ 
\frac{1}{2}(|0\rangle^{\otimes n}+|1\rangle^{\otimes n})
\]
if the measurement outcome was `$+1$' and
\[
\frac{1}{2}(-|0\rangle^{\otimes n}+|1\rangle^{\otimes n})
\]
if the outcome was `$-1$'.
In both cases we get a highly entangled state
showing maximal quantum uncertainty with respect to 
the observable $\overline{\sigma}_z$.

One might ask, whether
one can go beyond the  bound given by Theorem \ref{Main}
if one uses 
 additional measurements in between some steps of the quantum network.
It is easy to see that this possibility is already included in Theorem
\ref{Main}:
Every measurement procedure
which can be performed {\it directly} can be considered
as a {\it local} map which depends on the measurement outcome.
Intermediate local transformations and
  the preceding or the 
following bilocal transformation accessing the same site 
can be contracted to a single bilocal map. 
Then the additional local maps do not change the depth at all. 
Roughly speaking, we have found:
`measurements with constant complexity cannot produce that kind of highly entangled states which have large parameter $e_\rho$.'

The following theorem elucidates this result from another point of view
by showing that observables which are measurable (in the strong sense)
 with very low complexity
are almost compatible with the averaging ones, in the sense that their
spectral projections almost commute with the former ones:

\begin{Theorem}
Let $A$ be an  quantum network of depth $k$ consisting of unitary transformations.
 Let $c$ be an observable which
can be measured directly in the strong sense.
Then we have for every spectral projection $P$ of $c$:
\[
\|[\overline{a},A^*(P)]\|\leq \frac{2^k}{\sqrt{2n}}
\]
for every averaging observable $\overline{a}$.
\end{Theorem}

\begin{Proof}
Since $A^*$ is implemented by unitary transformations we have
\[
\|[\overline{a},A^*(P)]\|=\|[A(\overline{a}),P]\|\,.
\]
Set $c_i:=A(a_i)$. We find
\beq\label{Kommu}
\|\sum_i c_iP-P\sum_i c_i\|^2=\|(\sum_i c_iP-P\sum_ic_i)^2\|\,.
\eeq
We set $d_i:=c_i-Pc_iP$. Obviously $[d_i,P]=[c_i,P]$. Since $Pd_iP=0$
 we can write
expression (\ref{Kommu}) as
\begin{eqnarray}\label{Kommu2}
&&\|\sum_{ij}(d_iP-Pd_i)(d_jP-Pd_j)\| \nonumber 
\\ &=&\|(\sum_id_iP)(\sum_i d_iP)^\dagger
+(\sum_id_iP)^\dagger(\sum_i d_iP)\| \nonumber  \\
&\leq&2 \|\sum_iPd_i\|^2\leq 2\sum_{i,j}\|Pd_id_jP\|\, . 
\end{eqnarray}

Easy  calculation shows
\[
Pd_i=Pc_iP^\perp \,\, \hbox{ and }\,\, d_jP=P^\perp c_j P \,.
\]
Therefore we obtain
\[
\|[\sum_ic_i,P]\|^2\leq 2\sum_{i,j} \|Pc_iP^\perp c_j P\|
\]
Now we argue that the sum has to be taken over those pairs $(i,j)$ only which
have the property that the supports of $c_i$ and $c_j$ intersect:
By assumption $P$ has the form
\[
P=\otimes_i Q_i
\]
where every $Q_i$ is a projection on the i-th site.
The orthogonal projection $P^\perp$ can be split up into a direct sum in the 
following way:
For any binary word $b\in \{0,1\}^n$ define the projection
\[
R_b:=\otimes_{i\leq n}  S_i
\]
where $S_i=Q_i$ if the $i$-th digit of $b$ is `0' and
$S_i=Q_i^\perp$ otherwise.
Then $P^\perp$ can be written as:
\[
P^\perp=\oplus_{b\in \{0,1\}^n\setminus \{0\dots 0\}}\,\, R_b\,.
\] 
Let $r\in \{1,\dots,n\}$ be such that $b$ has the digit `1' at the site $r$.
 If the supports of $c_i$ and $c_j$ are disjoint, the site $r$ cannot  be
an element of both. Hence at least one of the two terms  
\[
Pc_iR_b \,\, \hbox{ and } \,\, R_bc_jP
\] 
vanishes, since $Q^\perp_r c_j Q_r=0$  if $r$ is not an element of the support
of $c_j$.

In the proof of Theorem \ref{Main} we have shown that for every 
$c_i$ there are 
at most $4^k$ operators $c_j$ such that their supports intersect.
Since $\|c_i\|\leq \|a_i\|\leq 1/2$ we have
\[
\|[A(\overline{a}),P]\| \leq \sqrt{\frac{1}{2n} 4^k}=\sqrt{\frac{1}{2n}}2^k\,.
\]
\end{Proof}

\section{Conclusions}

In agreement with our approach in \cite{JB} we introduced a parameter $e_\rho$
indicating that a many-particle state $\rho$ shows {\it quantum}
uncertainty on the macroscopic level. We showed that this class
of highly entangled states with large $e_\rho$ requires
a  quantum network with depth $\Theta (\log n)$.
In \cite{JB} we could give a set of  families of parallel hypersurfaces, where
 every family is parameterized by $r\in [-1,1]$.
For the hypersurfaces with increasing $|r|$ we could prove increasing
lower bounds for the depth of a quantum network preparing states lying
outside those pairs corresponding to $\pm r$. 
Since we consider states with large parameter $e_\rho$
as those one constituting the philosophical problem
of Schr\"odinger's Cat in the debate about {\it macrorealism}
we have proven that states which are prepared by a quantum network
of depth $O(1)$ are consistent with macrorealism.

In analogy, we investigated the depth of a network required for
measuring those observables which constitute the conflict with macrorealism
in the sense we had explained.
Here we obtained the same $\Theta (\log n)$ - bound as for the preparation
of states with large $e_\rho$.

Our results put strong restrictions on the set of entangled states
which can be obtained by networks with constant depth
including  intermediate 
measurements. Further understanding of `very-low-complexity entanglement'
 has to be left to the future. Since any tensor product structure
of a state for $n$ 
particles can already  be destroyed  by a quantum network of depth 2,
this seems to be non-trivial at all.
\end{multicols}

\section*{Acknowledgments}

Many thanks to M. Grassl, M. R\"otteler, and P. Wocjan for 
useful discussions. This work was partially supported by  the
project `AQUA' of the `Deutsche Forschungsgemeinschaft'.

\end{document}